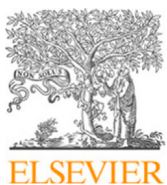
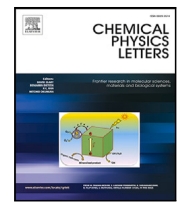

# Post-CCSD(T) corrections in the S66 noncovalent interactions benchmark

Emmanouil Semidalas [a], A. Daniel Boese [b], Jan M.L. Martin [a],*

[a] Department of Molecular Chemistry and Materials Science, Weizmann Institute of Science, 7610001 Reḥovot, Israel
[b] Department of Chemistry, University of Graz, Heinrichstrasse 28/IV, 8010 Graz, Austria

## HIGHLIGHTS

- Post-CCSD(T) corrections for most of the S66 noncovalent interactions benchmark
- For most systems, higher-order triples and connected quadruples nicely cancel
- For $\pi$-stacking systems, however, higher-order triples are much more important
- Hence, the cancellation breaks down, and CCSD(T) tends to overbind
- This corroborates recent claims based on diffusion Monte Carlo
- Simple estimation formulas for post-CCSD(T) contributions are proposed.

## GRAPHICAL ABSTRACT

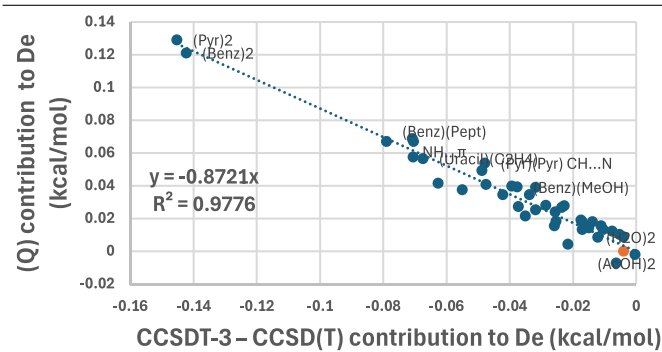

## ARTICLE INFO

## ABSTRACT

For noncovalent interactions, it is generally assumed that CCSD(T) approaches the exact solution within the basis set. For most of the S66 benchmark, we present CCSDT and CCSDT(Q) corrections with a DZP basis set. For hydrogen bonds, pure London, and mixed-influence complexes, CCSD(T) benefits from error cancellation between (repulsive) higher-order triples, $T_3 - (T)$, and (attractive) connected quadruples, (Q). For $\pi$-stacking complexes, this cancellation starts breaking down and CCSD(T) overbinds; CCSD(T)$_A$ corrects the problem at the expense of London complexes. Simple two- or three-parameter models predict CCSDT(Q)–CCSD(T) differences to 0.01 kcal mol$^{-1}$ RMS, requiring no calculations with steeper scaling than $O(N^7)$.

## 1. Introduction

The importance of noncovalent interactions in supramolecular chemistry, molecular recognition, and the structure of condensed matter hardly needs to be reiterated. As individual interactions are difficult to access experimentally, accurate ab initio wavefunction theory (WFT) calculations have become the primary source of training and validation data for more approximate approaches. (For recent reviews, see Refs. [1,2]; Ref. [3] focuses specifically on biomolecules and Ref. [4] on aromatic systems.)

CCSD(T) [5] near the complete basis set limit has become 'the gold standard' for such calculations, as it is well known (e.g., Refs. [6–8]) that it benefits from an error compensation between (typically antibonding) higher-order triples $T_3 - (T)$ and (bonding) connected quadruples (Q), as illustrated in Fig. 1.

Very recently, however, al-Hamdani et al. [11] showed that for ever larger $\pi$-stacking complexes, an increasingly larger gap opens between local natural orbital CCSD(T) [12,13] and fixed node-diffusion Monte Carlo (FN-DMC) [14] results, which purportedly are more accurate (devoid of systematic error) even as they suffer from lower precision (greater stochastic error). Grüneis and coworkers [15] attributed the discrepancy to CCSD(T) overshooting the effect of triple excitations in systems with large polarizabilities, which they call the 'infrared catastrophe' [16]. It does appear from Fig. 1 in Ref. [15] that their CCSD(cT) approach [16], which can be seen as a noniterative approximation to CCSDT-2 [17], addresses the problem.

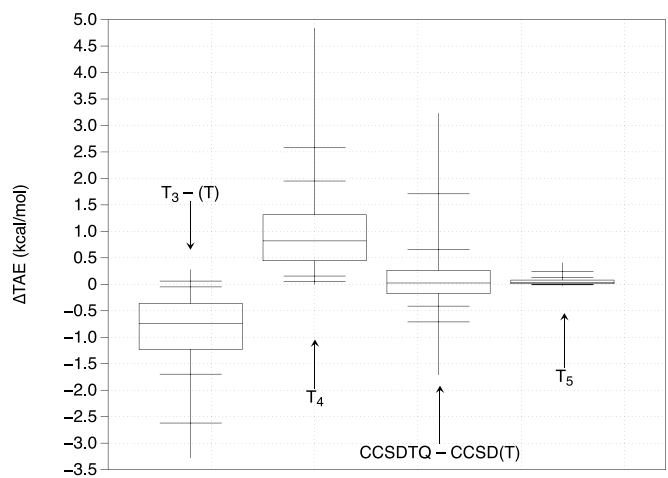

**Fig. 1.** Box-and-whiskers plot of the total atomization energy contributions of higher-order corrections in the W4-17 dataset [9]. Outer fences encompass 95% of the set, inner fences 80%, boxes 50%. Vertical lines span from population minimum to maximum. Copyright Springer Nature.
*Source:* Reprinted from Figure 2 in Ref. [10].

But this leaves the reader with as many questions as answers. What about CCSDT, i.e., coupled cluster with all singles, doubles, and fully iterative triples? And would connected quadruples (Q) again disrupt agreement with FN-DMC? Moreover, how representative is this?

Owing to the very steep computational cost scaling of approaches such as CCSDT, $O(n_{occ}^3 N_{virt}^5)$ (where $n_{occ}$ and $N_{virt}$ are the numbers of occupied and virtual orbitals, respectively), and CCSDT(Q), $O(n_{occ}^4 N_{virt}^5)$, post-CCSD(T) studies of noncovalent complexes are fairly scant [15,18–20], and mostly limited to very small monomers. (Some exceptions that come to mind are Refs. [15,21,22]). The aim of this paper is to assess different approximations which maintain the original CCSD(T) CPU time scaling (such as the CCSD(cT) approach or the CCSDT-2 and CCSDT-3 approximations to CCSDT), and/or by fitting different empirical corrections to match CCSDT(Q) as closely as possible. We will also consider the CCSD(T)$_\Lambda$ [23–26] approach, which treats the triples as a perturbation to CCSD. (The CCSD(T) equations can be derived [27] by approximating the $\Lambda$ vector as the transpose of the doubles amplitudes vector $\mathbf{T}_2$.)

The S66 noncovalent interactions benchmark of Hobza and coworkers [28] consists of building blocks of biomolecules in different interactions (for example, uracil dimer in both Watson–Crick and stacked configurations) and would seem to be a good test case. We have previously [29] revised the CCSD(T) basis set limits for this dataset; in the present work, we shall attempt to evaluate post-CCSD(T) corrections for at least a representative subset of S66.

## 2. Computational details

All electronic structure calculations in this work were carried out using a development version of the CFOUR [30] electronic structure program system, run on the CHEMFARM cluster of the Faculty of Chemistry at Weizmann. The methods beyond standard CCSD(T) [5] include fully iterative CCSDT [31], the CCSDT-1b and CCSDT-3 approximations thereto [17,32], CCSDT(Q) [33], and CCSD(T)$_\Lambda$ [23–26]. (We note, e.g., Ref. [34] that the difference between CCSDT-2 and CCSDT-3 only sets in at sixth order in perturbation theory. It results from the effect on the $T_3$ amplitudes equation of single excitations, which is neglected in CCSDT-2 but included in CCSDT-3.)

The basis sets used are of the correlation consistent [35–38] family. The shorthand cc-pVDZ(d,s) refers to at most $d$ and $s$ functions, respectively, on nonhydrogen and hydrogen atoms (the full cc-pVDZ basis set would correspond to cc-pVDZ(d,p), and the smallest basis set of this series without any polarized functions to cc-pVDZ(p,s).)

Reference geometries from the S66 [28] dataset were initially taken as-is from the BEGDB database [39]. These geometries were originally optimized at the BSSE corrected MP2/cc-pVTZ level without any symmetry, and typically only have $C_1$ symmetry, but in some cases, they are very close to structures with $C_s$, $C_{2v}$, or $C_{2h}$ symmetries. In several cases (such as **24** benzene...benzene and **25** pyridine...pyridine parallel-displaced, as well as **34** pentane...pentane), we resorted to treating the symmetrized structure in order to be able to bring the calculations to completion at all. Even so, they took over a month each on compute nodes with 52 cores and 768 GB RAM.

No BSSE corrections were applied, as it was shown in a separate study [40] on S66 that BSSE contributions to post-CCSD(T) correlation contributions are negligible.

## 3. Results and discussion

Relevant components of dissociation energies ($D_e$ values) are presented in Table 1, while Table 2 offers performance statistics broken down into the four primary categories from Refs. [28,41]: hydrogen bonds (systems **1–23**), $\pi$-stacking (systems **24–33**), pure London dispersion (systems **34–46**), and mixed-influence complexes. CCSDT-3–CCSDT and CCSDT(Q)-CCSD(T) differences larger than 0.05 kcal mol$^{-1}$ are found for AcOH...AcOH, AcOH...uracil, benzene and pyridine dimers, as well as for all the aromatic–aromatic $\pi$ stacks, systems **24**...**29**.

As expected, higher-order triples are antibonding, most pronouncedly so for $\pi$ stacks.

CCSDT-3 appears to recover the lion's share of the higher-order triples effects: the RMS difference (RMSD) with CCSDT drops from 0.08 kcal mol$^{-1}$ for CCSD(T) to 0.029 kcal mol$^{-1}$ for CCSDT-3. The most pronounced improvement is seen for $\pi$ stacks, followed by mixed-influence systems. A detailed perturbation theoretical analysis of the differences between various approximate triple-excitations methods has been published by Cremer and coworkers; [34] specifically,

$$E[\text{CCSDT-3}] - E[\text{CCSD(T)}] = E_{TQ}^{[5]} + O(\lambda^6) \tag{1}$$

where $\lambda$ is the perturbation parameter and $E_{TQ}^{[5]}$ is the fifth-order triples-disconnected quadruples interaction term.

The CCSDT – CCSDT-3 difference corresponds to the triples–triples interaction term, which again starts at fifth order:

$$E[\text{CCSDT}] - E[\text{CCSDT-3}] = E_{TT}^{[5]} + O(\lambda^6) \tag{2}$$

$E_{TQ}^{[5]}$ appears to be antibonding across the board. The fifth-order connected triples–triples term $E_{TT}^{[5]}$, on the other hand, is found to be somewhat antibonding for $\pi$ stacks, slightly bonding for London complexes, and almost nil for the two remaining categories.

Grüneis et al. [15] in the Supporting Information of their paper, found that the difference between CCSD(T) and their CCSD(cT) for a set of nine $\pi$-stacks is surprisingly well described by the following model:

$$\Delta E[(cT)] \approx \frac{\Delta E[(T)]}{a_1 + a_2 \frac{\Delta E_2}{\Delta E_{\text{corr,CCSD}}}} \tag{3}$$

This equation can be slightly rearranged as follows, with $\gamma = 1/a_1$ and $\alpha = a_2/a_1$:

$$\Delta E[cT] - \Delta E[(T)] \approx \Delta E[(T)] \left( \frac{\gamma}{1 + \alpha \left( \frac{\Delta E_2}{\Delta E_{\text{corr,CCSD}}} - 1 \right)} - 1 \right) \tag{4}$$

**Table 1**
Convergence along the coupled cluster series of S66 $D_e$ (kcal mol$^{-1}$) with the cc-pVDZ(d,s) basis set.

| $D_e$ components | HF | MP2 –HF | CCSD –HF | (T) | CCSD(T)$_A$ –CCSD(T) | CCSDT-1b –CCSD(T) | CCSDT-3 –CCSD(T) | CCSDT– CCSDT-3 | (Q) | CCSDT(Q) –CCSD(T) | CCSDT(Q) –CCSDT-3 |
|---|---|---|---|---|---|---|---|---|---|---|---|
| 1 Water…Water | 6.102 | 1.342 | 0.836 | 0.154 | −0.002 | 0.015 | −0.004 | 0.004 | 0.009 | 0.008 | 0.012 |
| 2 Water…MeOH | 6.077 | 1.634 | 1.096 | 0.188 | −0.004 | 0.019 | −0.008 | 0.002 | 0.012 | 0.007 | 0.015 |
| 3 Water…MeNH$_2$ | 6.744 | 1.957 | 1.246 | 0.256 | −0.012 | 0.017 | −0.017 | 0.010 | 0.018 | 0.011 | 0.028 |
| 4 Water…peptide | 7.831 | 1.398 | 0.932 | 0.199 | −0.014 | 0.007 | −0.022 | 0.005 | 0.004 | −0.012 | 0.009 |
| 5 MeOH…MeOH | 6.285 | 2.185 | 1.536 | 0.276 | −0.006 | 0.028 | −0.011 | 0.005 | 0.016 | 0.009 | 0.020 |
| 6 MeOH…MeNH$_2$ | 6.985 | 3.058 | 2.136 | 0.423 | −0.017 | 0.032 | −0.024 | 0.020 | 0.027 | 0.023 | 0.047 |
| 7 MeOH…peptide | 8.218 | 2.679 | 1.947 | 0.402 | −0.017 | 0.037 | −0.026 | 0.007 | 0.016 | −0.004 | 0.022 |
| 8 MeOH…water | 6.198 | 1.712 | 1.130 | 0.215 | −0.004 | 0.021 | −0.005 | 0.006 | 0.010 | 0.011 | 0.017 |
| 9 MeNH$_2$…MeOH | 2.882 | 1.914 | 1.563 | 0.234 | −0.005 | 0.021 | −0.011 | 0.011 | 0.013 | 0.014 | 0.024 |
| 10 MeNH$_2$…MeNH$_2$ | 2.915 | 2.823 | 2.283 | 0.337 | −0.016 | 0.018 | −0.026 | 0.021 | 0.024 | 0.020 | 0.045 |
| 11 MeNH$_2$…peptide | 3.904 | 3.214 | 2.601 | 0.422 | −0.019 | 0.030 | −0.035 | 0.017 | 0.022 | 0.003 | 0.038 |
| 12 MeNH$_2$…water | 6.738 | 2.257 | 1.486 | 0.271 | −0.012 | 0.018 | −0.017 | 0.011 | 0.019 | 0.013 | 0.030 |
| 13 Peptide…MeOH | 6.019 | 2.970 | 2.281 | 0.361 | −0.013 | 0.030 | −0.025 | 0.008 | 0.018 | 0.001 | 0.026 |
| 14 Peptide…MeNH$_2$ | 6.175 | 3.811 | 2.899 | 0.517 | −0.028 | 0.033 | −0.042 | 0.023 | 0.034 | 0.015 | 0.057 |
| 15 Peptide…peptide | 7.174 | 3.396 | 2.616 | 0.527 | −0.034 | 0.037 | −0.055 | 0.008 | | | |
| 16 Peptide…water | 5.985 | 1.943 | 1.451 | 0.238 | −0.008 | 0.017 | −0.012 | 0.009 | 0.009 | 0.005 | 0.017 |
| 17 Uracil…uracil (BP) | 15.275 | 2.839 | 1.682 | 0.811 | −0.075 | 0.071 | −0.096 | | | | |
| 18 Water…pyridine | 5.947 | 1.879 | 1.148 | 0.265 | −0.011 | 0.022 | −0.023 | −0.002 | 0.028 | 0.003 | 0.025 |
| 19 MeOH…pyridine | 6.313 | 2.889 | 1.871 | 0.426 | −0.016 | 0.041 | −0.032 | −0.001 | 0.039 | 0.006 | 0.037 |
| 20 AcOH…AcOH | 18.464 | 2.158 | 0.859 | 0.699 | −0.033 | 0.117 | −0.006 | −0.042 | −0.007 | −0.055 | −0.049 |
| 21 AcNH$_2$…AcNH$_2$ | 15.016 | 2.647 | 1.818 | 0.632 | −0.058 | 0.053 | −0.051 | 0.011 | 0.015 | −0.025 | 0.025 |
| 22 AcOH…uracil | 18.293 | 2.259 | 1.090 | 0.717 | −0.067 | 0.060 | −0.071 | −0.028 | 0.013 | −0.086 | −0.015 |
| 23 AcNH$_2$…uracil | 17.304 | 2.698 | 1.608 | 0.752 | −0.084 | 0.038 | −0.099 | −0.009 | | | |
| 24 Benzene…benzene ($\pi…\pi$) | −4.027 | 7.895 | 5.110 | 0.980 | −0.064 | 0.085 | −0.142 | −0.064 | 0.121 | −0.085 | 0.057 |
| 25 Pyridine…pyridine ($\pi…\pi$) | −3.091 | 8.459 | 5.402 | 1.040 | −0.064 | 0.102 | −0.145 | −0.082 | 0.129 | −0.099 | 0.047 |
| 26 Uracil…uracil ($\pi…\pi$) | 2.649 | 9.096 | 6.606 | 1.414 | −0.047 | 0.196 | −0.120 | −0.068 | | | |
| 27 Benzene…pyridine ($\pi…\pi$) | −3.449 | 8.144 | 5.248 | 1.009 | −0.064 | 0.093 | −0.144 | −0.073 | | | |
| 28 Benzene…uracil ($\pi…\pi$) | −1.838 | 8.929 | 6.197 | 1.204 | −0.059 | 0.134 | −0.144 | −0.081 | | | |
| 29 Pyridine…uracil ($\pi…\pi$) | −0.463 | 8.644 | 5.962 | 1.175 | −0.063 | 0.126 | −0.149 | −0.082 | | | |
| 30 Benzene…ethene | −2.651 | 4.118 | 2.862 | 0.496 | −0.037 | 0.024 | −0.070 | −0.010 | 0.067 | −0.013 | 0.057 |
| 31 Uracil…ethene | −0.291 | 3.792 | 2.808 | 0.507 | −0.035 | 0.029 | −0.067 | −0.007 | 0.057 | −0.018 | 0.050 |
| 32 Uracil…ethyne | 0.502 | 3.221 | 2.276 | 0.453 | −0.036 | 0.024 | −0.071 | −0.015 | 0.057 | −0.029 | 0.042 |
| 33 Pyridine…ethene | −2.214 | 4.305 | 2.978 | 0.517 | −0.037 | 0.029 | −0.071 | −0.014 | 0.069 | −0.016 | 0.055 |
| 34 Pentane…pentane | −2.476 | 5.639 | 4.926 | 0.678 | −0.036 | 0.030 | −0.063 | 0.047 | 0.041 | 0.026 | 0.088 |
| 35 Neopentane…pentane | −1.375 | 4.023 | 3.573 | 0.501 | −0.027 | 0.022 | −0.046 | 0.035 | | | |
| 36 Neopentane…neopentane | −0.590 | 2.887 | 2.608 | 0.366 | −0.020 | 0.015 | −0.033 | 0.029 | | | |
| 37 Cyclopentane…neopentane | −1.575 | 4.055 | 3.554 | 0.508 | −0.027 | 0.024 | −0.047 | 0.033 | | | |
| 38 Cyclopentane…cyclopentane | −1.791 | 4.634 | 4.005 | 0.594 | −0.032 | 0.030 | −0.055 | 0.036 | 0.037 | 0.019 | 0.074 |
| 39 Benzene…cyclopentane | −1.965 | 6.397 | 4.737 | 0.806 | −0.048 | 0.054 | −0.095 | −0.006 | | | |
| 40 Benzene…neopentane | −1.285 | 4.966 | 3.757 | 0.634 | −0.039 | 0.039 | −0.073 | 0.000 | | | |
| 41 Uracil…pentane | −1.202 | 7.148 | 5.752 | 0.990 | −0.045 | 0.085 | −0.094 | | | | |
| 42 Uracil…cyclopentane | −1.129 | 6.337 | 5.068 | 0.884 | −0.038 | 0.078 | −0.084 | | | | |
| 43 Uracil…neopentane | −0.346 | 4.719 | 3.816 | 0.661 | −0.033 | 0.053 | −0.066 | | | | |
| 44 Ethene…pentane | −0.857 | 2.682 | 2.276 | 0.324 | −0.020 | 0.012 | −0.032 | 0.021 | 0.025 | 0.014 | 0.046 |
| 45 Ethyne…pentane | −0.933 | 2.507 | 1.924 | 0.298 | −0.020 | 0.012 | −0.034 | 0.009 | 0.034 | 0.010 | 0.044 |
| 46 Peptide…pentane | −0.471 | 5.529 | 4.676 | 0.713 | −0.030 | 0.053 | −0.060 | 0.029 | | | |
| 47 Benzene…benzene (TS) | −0.870 | 4.573 | 3.211 | 0.581 | −0.033 | 0.049 | −0.070 | −0.025 | | | |
| 48 Pyridine…pyridine (TS) | 0.143 | 4.608 | 3.228 | 0.590 | −0.030 | 0.057 | −0.068 | −0.028 | | | |
| 49 Benzene…pyridine (TS) | −0.341 | 4.468 | 3.135 | 0.571 | −0.030 | 0.050 | −0.068 | −0.028 | | | |
| 50 Benzene…ethene (CH–$\pi$) | 0.807 | 2.728 | 1.839 | 0.341 | −0.017 | 0.026 | −0.039 | −0.014 | 0.040 | −0.013 | 0.026 |
| 51 Ethyne…ethyne (TS) | 0.837 | 0.849 | 0.568 | 0.138 | −0.011 | 0.005 | −0.015 | 0.002 | 0.014 | 0.001 | 0.016 |
| 52 Benzene…AcOH (OH–$\pi$) | 2.521 | 3.423 | 2.379 | 0.478 | −0.018 | 0.057 | −0.035 | −0.028 | | | |
| 53 Benzene…AcNH$_2$ (NH–$\pi$) | 2.523 | 3.022 | 2.209 | 0.433 | −0.023 | 0.043 | −0.037 | −0.013 | | | |
| 54 Benzene…water (OH–$\pi$) | 2.289 | 1.656 | 1.126 | 0.207 | −0.008 | 0.019 | −0.014 | −0.009 | 0.018 | −0.005 | 0.009 |
| 55 Benzene…MeOH (OH–$\pi$) | 1.450 | 3.569 | 2.566 | 0.459 | −0.021 | 0.039 | −0.038 | −0.010 | 0.039 | −0.009 | 0.029 |
| 56 Benzene…MeNH$_2$ (NH–$\pi$) | 0.030 | 3.632 | 2.688 | 0.450 | −0.026 | 0.029 | −0.048 | −0.003 | 0.041 | −0.010 | 0.038 |
| 57 Benzene…peptide (NH–$\pi$) | 0.839 | 5.493 | 3.951 | 0.709 | −0.042 | 0.054 | −0.079 | −0.019 | 0.067 | −0.031 | 0.048 |
| 58 Pyridine…pyridine (CH–N) | 2.702 | 3.177 | 2.415 | 0.459 | −0.019 | 0.053 | −0.048 | −0.012 | 0.054 | −0.007 | 0.041 |
| 59 Ethyne…water (CH–O) | 3.895 | 1.047 | 0.745 | 0.136 | 0.001 | 0.017 | 0.000 | 0.004 | −0.002 | 0.001 | 0.002 |
| 60 Ethyne…AcOH (OH–$\pi$) | 4.024 | 1.536 | 0.947 | 0.313 | −0.016 | 0.036 | −0.017 | −0.009 | 0.013 | −0.013 | 0.004 |
| 61 Pentane…AcOH | 0.033 | 3.734 | 3.150 | 0.487 | −0.017 | 0.041 | −0.036 | 0.018 | | | |
| 62 Pentane…AcNH$_2$ | 0.393 | 4.444 | 3.730 | 0.601 | −0.025 | 0.047 | −0.049 | 0.024 | | | |
| 63 Benzene…AcOH | 0.106 | 4.327 | 3.167 | 0.552 | −0.026 | 0.048 | −0.060 | −0.020 | | | |
| 64 Peptide…ethene | 1.049 | 2.770 | 2.221 | 0.377 | −0.019 | 0.025 | −0.037 | 0.012 | 0.027 | 0.002 | 0.039 |
| 65 Pyridine…ethyne | 3.747 | 1.778 | 1.148 | 0.277 | −0.010 | 0.030 | −0.029 | −0.004 | 0.028 | −0.005 | 0.024 |
| 66 MeNH$_2$…Pyridine | 0.910 | 3.788 | 2.798 | 0.470 | −0.025 | 0.036 | −0.049 | −0.002 | 0.049 | −0.001 | 0.047 |

No BSSE correction was applied.

in which $\gamma$ is an optional scaling factor and $\alpha$ can be interpreted as a regularization parameter. If the ratio of the MP2 correlation energy $E_2$ to the CCSD correlation energy $E_{\mathrm{corr,CCSD}}$ is near unity, then the denominator will be largely unaffected; if, however, MP2 grossly overestimates the CCSD correlation energy because of near-degeneracy, then the denominator will 'throttle' the quasiperturbative triples to

**Table 2**
Breakdown of average and RMS post-CCSD(T) contributions (kcal mol$^{-1}$) into the four major subcategories of the S66 dataset.

|  | (T)$_A$ –(T) | CCSDT-1b– CCSD(T) | CCSDT-3– CCSD(T) | CCSDT– CCSDT-3 | CCSDT– CCSD(T) | (Q) | CCSDT(Q) –CCSD(T) | CCSDT(Q) –CCSDT-3 | CCSDT(Q) –CCSD(T)$_A$ |
|---|---|---|---|---|---|---|---|---|---|
|  | Average |  |  |  |  |  |  |  |  |
| H-bond | −0.024 | 0.034 | −0.031 | 0.004 | −0.024 | 0.017 | −0.002 | 0.022 | 0.016 |
| $\pi$-stack | −0.051 | 0.084 | −0.112 | −0.050 | −0.162 | 0.083 | −0.043 | 0.051 | 0.002 |
| pure London | −0.032 | 0.039 | −0.060 | 0.023 | −0.030 | 0.035 | 0.017 | 0.063 | 0.044 |
| mixed influence | −0.021 | 0.038 | −0.042 | −0.008 | −0.050 | 0.032 | −0.006 | 0.027 | 0.010 |
| All S66 | −0.029 | 0.044 | −0.052 | −0.005 | −0.056 | 0.032 | −0.006 | 0.031 | 0.015 |
|  | RMSD |  |  |  |  |  |  |  |  |
| H-bond | 0.034 | 0.041 | 0.041 | 0.015 | 0.038 | 0.020 | 0.026 | 0.031 | 0.023 |
| $\pi$-stack | 0.052 | 0.101 | 0.118 | 0.059 | 0.175 | 0.088 | 0.056 | 0.052 | 0.022 |
| pure London | 0.033 | 0.045 | 0.064 | 0.029 | 0.042 | 0.035 | 0.018 | 0.066 | 0.046 |
| mixed influence | 0.023 | 0.041 | 0.046 | 0.017 | 0.058 | 0.037 | 0.011 | 0.031 | 0.012 |
| All S66 | 0.034 | 0.055 | 0.064 | 0.029 | 0.083 | 0.043 | 0.029 | 0.039 | 0.024 |

The cc-pVDZ(d,s) basis set was used throughout.

compensate, with $\beta$ acting as the throttle lever.

We note that nonlinear corrections like Eqs. (3), (4) as well as Eqs. (9), (11), (12) below, are not size-consistent, i.e., they will yield a different result when applied to the correlation energies of the different species — in this case, the ratio $D_e[(T)]/D_e[CCSD_{corr}]$ will be much larger than the individual $E(T)/E_{corr}[CCSD]$ ratios.

Eq. (3), with parameters $a_1 = 0.9417$ and $a_2 = 0.1442$ fitted in this work, performs fairly poorly for $T_3 - (T)$, its RMSD of 0.035 kcal mol$^{-1}$ being almost half of the actual effect, 0.083 kcal/mol RMS. In contrast, CCSDT-3–CCSD(T) is modeled moderately well at RMSD = 0.016 kcal mol$^{-1}$— one-quarter of the actual effect, meaning the model recovers about three-quarters of the variation in it. However, the fitted model parameters $a_1 = 1.140$, $a_2 = -0.016$ suggest that a simple scaling of (T) would work equally well, and indeed the same RMSD = 0.016 kcal mol$^{-1}$ is obtained by 0.895 $\Delta E[(T)]$ for all of $\Delta E[CCSDT-3] - \Delta E[CCSD]$, which works out to $-0.105 E[(T)]$ for CCSDT-3 – CCSD(T) by itself.

Surprisingly, however, it turns out CCSDT–CCSD(T) can be modeled to RMSD = 0.009 kcal/mol by also considering the CCSDT-1b method, which can be seen as an iterative counterpart to CCSD(T). That is:

$$E[T_3 - (T)] \approx c_1 (E[CCSDT-3] - E[CCSD(T)])$$
$$+ c_2 (E[CCSDT-1b] - E[CCSD(T)])$$
$$+ (1 - c_1 - c_2) E[(T)] \quad (5)$$

with $c_1 = 2.130$, $c_2 = -1.345$.

This equation, being linear, is size-extensive to boot, at the expense of requiring an additional $O(n_{occ}^3 N_{virt}^4)$ calculation step, but still having no steeper scaling than $O(n_{occ}^3 N_{virt}^4)$.

CCSDT(Q) is exact to fifth order: the additional terms that (Q) introduces beyond CCSDT are

$$E[CCSDT(Q)] - E[CCSDT] = E_{QQ}^{[5]} + E_{QT}^{[5]} + O(\lambda^6) \quad (6)$$

where $E_{QQ}^{[5]}$ is the fifth-order connected quadruples term, and it should be noted that the disconnected quadruples–triples term $E_{QT}^{[5]}$ is the Hermitian conjugate of $E_{TQ}^{[5]}$. Hence, for real orbitals, CCSD(T) differs from CCSDT(Q) by three fifth-order terms:

$$E[CCSDT(Q)] - E[CCSD(T)] = E_{QQ}^{[5]} + 2E_{TQ}^{[5]} + E_{TT}^{[5]} + O(\lambda^6) \quad (7)$$

Now (Q) is attractive across the board, by 0.043 kcal mol$^{-1}$ RMS, and unsurprisingly its greatest importance by far is for the $\pi$ stacks, as it compensates for the aforementioned strong effects found in the higher-order triples terms.

An Eq. (3) type model performs somewhat indifferently when fitted to the CCSDT(Q)-CCSDT differences, yielding an RMS = 0.017 kcal mol$^{-1}$ with $a_1 = 0.847$, $a_2 = 0.052$. We could instead carry out a form of geometric extrapolation (a special case of Padé approximants [42]), in which we assume that $E_{corr,CCSD}$, $E(T)$, and $E(Q)$ converge in a geometric series. The latter implies, if we define the geometric decay ratio $r = E(T)/E_{corr,CCSD}$, that $E(Q) \approx r.E(T) = E(T)^2/E_{corr,CCSD}$. Summation to the full-CI limit would then lead to

$$E_{FCI} - E_{corr,CCSD} \approx E(T) \times (1 + r + r^2 + r^3 + \cdots) = E(T)/(1 - r) \quad (8)$$

If we introduce two adjustable parameters, we obtain

$$\Delta E(Q) \approx \Delta E(T) \times \left( \frac{\alpha}{1 - \beta \frac{E(T)}{\Delta E_{corr}[CCSD]}} - 1 \right) \quad (9)$$

which for $\alpha = 1.127$ and $\beta = -0.168$ has a somewhat lower RMSD = 0.014 kcal mol$^{-1}$. Scaling the actual calculated $\Delta E[(Q)]$ obtained in the *unpolarized* cc-pVDZ(p,s) basis set by 1.267 yields RMSD = 0.011 kcal mol$^{-1}$, but still requires pretty costly calculations for larger systems.

In fact, simple scaling of the CCSDT-3 – CCSD(T) difference

$$\Delta E(Q) \approx -b_1 (\Delta E[CCSDT-3] - \Delta E[CCSD(T)]) \quad (10)$$

with $b_1 = 0.825$ achieves RMSD = 0.011 kcal mol$^{-1}$. (Actually, using the CCSDT–CCSD(T) difference scaled by the optimized factor $-0.626$ for the same purpose works less well — and CCSDT-3, which scales as $O(n_{occ}^3 N_{virt}^4)$, is clearly a more economical approach than CCSDT at $O(n_{occ}^3 N_{virt}^5)$.) Upon deleting the outlier points **21** and **22** RMSD can be improved to 0.0064 kcal/mol with $b_1 = 0.872$ (see Table of Contents graphic). Another way to look at this result is that most of the CCSDT(Q)–CCSD(T) difference will reflect CCSDT–CCSDT-3, since the two other components nearly cancel each other on average.

Now let us consider the post-CCSD(T) correction as a whole. Both CCSD(T) and CCSD(T)$_A$ benefit from error compensation here, with RMSDs of 0.026 and 0.024, respectively. However, while CCSD(T)$_A$ clearly performs best for $\pi$ stacks, it somewhat 'drops the ball' for pure London dispersion, while the opposite is true for CCSD(T).

A combination Grüneis and $E_5^{[TQ]}$ scaling model of the form

$$\Delta E[postCCSD(T)] \approx \Delta E(T) \left[ \frac{1}{a_1 + a_2 \frac{\Delta E_2}{\Delta E_{corr}[CCSD]}} - 1 \right]$$
$$+ a_3 (\Delta E[CCSDT-3] - \Delta E[CCSD(T)]) \quad (11)$$

reaches an RMSD = 0.009 kcal mol$^{-1}$ with $a_1 = 0.730$, $a_2 = 0.140$, and $a_3 = 0.968$. (If we substitute the still less expensive CCSD(T)$_A$, which has about twice the cost of CCSD(T), for CCSDT-3, the RMSD rises to 0.015 kcal mol$^{-1}$, with the associated parameters being 0.817, 0.090, and 1.448, respectively.) $a_3$ in Eq. (11) is close enough to unity that it immediately prompts the question, what would happen to RMSD if we just set $a_3 = 1$ and optimize only the two remaining parameters. The result is equivalent in quality: RMSD = 0.009 kcal mol$^{-1}$, $a_1 = 0.727$, $a_2 = 0.140$, $a_3 = 1$. (This, in fact, amounts to a Grüneis-type fit of CCSDT(Q) - CCSDT-3.)

Furthermore, a similar combined expression incorporating Eq. (9)

**Table 3**
Dependence of parameters in Eq. (12) on the basis set.

|  | cc-pVDZ(d,s) | cc-pVDZ(d,p) | cc-pVTZ(f,p) | haVDZ(d,p) | haVTZ(f,d) |
| --- | --- | --- | --- | --- | --- |
| $\alpha$ | 1.139 | 1.196 | 1.238 | 1.161 | 1.267 |
| $\beta$ | −0.273 | −0.571 | −0.824 | −0.504 | −1.009 |
| RMSD (kcal mol$^{-1}$) | 0.011 | 0.011 | 0.010 | 0.011 | 0.010 |

$a_3 = 1$ throughout. The notation haV$n$Z stands for cc-pV$n$Z on hydrogen combined with aug-cc-pV$n$Z on the remaining elements.

**Table 4**
Post-CCSD(T) contributions (kcal mol$^{-1}$) for nine $\pi$ stacks compared with earlier FN-DMC data.

|  | Ref. [15] | Refs. [11–15] | Obtained in present work, cc-pVDZ(d,s) basis | | | | | |
| --- | --- | --- | --- | --- | --- | --- | --- | --- |
|  | CCSD(cT) | FN-DMC– | CCSDT-3 | CCSDT | CCSDT(Q) | CCSDT(Q)-CCSD(T) | | |
|  | −CCSD(T) | CCSD(T)/CBS | −CCSD(T) | −CCSD(T) | −CCSDT | Eq. (13) | Eq. (10)+CCSDT | actual |
| 24 Benzene…benzene ($\pi\ldots\pi$) | −0.21 | −0.32 ± 0.12 | −0.142 | −0.206 | 0.121 | (−0.087) | (−0.089) | −0.085 |
| 25 Pyridine…pyridine ($\pi\ldots\pi$) | −0.23 | −0.26 ± 0.20 | −0.145 | −0.228 | 0.129 | (−0.105) | (−0.108) | −0.099 |
| 26 Uracil…uracil ($\pi\ldots\pi$) | −0.30 | −0.33 ± 0.16 | −0.120 | −0.188 | (0.10) | (−0.144) | (−0.089) | |
| 27 Pyridine…pyridine ($\pi\ldots\pi$) | −0.22 | −0.28 ± 0.16 | −0.144 | −0.217 | (0.12) | (−0.095) | (−0.098) | |
| 28 Benzene…uracil ($\pi\ldots\pi$) | −0.27 | −0.45 ± 0.18 | −0.144 | −0.225 | (0.12) | (−0.118) | (−0.106) | |
| 29 Pyridine…uracil ($\pi\ldots\pi$) | −0.27 | −0.29 ± 0.18 | −0.149 | −0.232 | (0.13) | (−0.117) | (−0.109) | |
| 47 Benzene…benzene (TS) | −0.12 | −0.11 ± 0.12 | −0.070 | −0.095 | (0.06) | (−0.034) | (−0.038) | |
| 48 Pyridine…pyridine (TS) | −0.12 | −0.05 ± 0.20 | −0.068 | −0.096 | (0.06) | (−0.042) | (−0.040) | |
| 49 Benzene…pyridine (TS) | −0.12 | −0.20 ± 0.16 | −0.068 | −0.096 | (0.06) | (−0.036) | (−0.040) | |

CBS = complete basis set (limit). Extrapolated CCSD(T)/CBS values taken from Table S I in Ref. [15], FN-DMC values with uncertainties (2$\sigma$) from Table 1 in Ref. [11]. These latter authors consider the difference between localized natural orbital coupled cluster, LNO-CCSD(T)/CBS and FN-DMC to be different from statistical zero only in cases **24** and **28**, by −0.1 kcal mol$^{-1}$ in both cases. Model estimates for (Q) and CCSDT(Q)-CCSD(T) are given in parentheses.

$$\Delta E[\text{postCCSD(T)}] \approx \Delta E(T) \times \left( \frac{\alpha}{1 - \beta \frac{\Delta E(T)}{\Delta E_{\text{corr}}[\text{CCSD}]}} - 1 \right) + a_3(\Delta E[\text{CCSDT-3}] - \Delta E[\text{CCSD(T)}]) \quad (12)$$

yields RMSD = 0.011 kcal mol$^{-1}$ for $\alpha = 1.167$, $\beta = -0.294$, and $a_3 = 1.212$. If we fix $a_3 = 1$ — which effectively makes the first term of Eq. (12) a geometric-style fit of CCSDT(Q) - CCSDT-3 — then RMSD is unchanged to three decimal places for $\alpha = 1.139$, $\beta = -0.273$.

Once again, if we additionally have CCSDT-1b available, this permits a size-consistent alternative:

$$E[\text{postCCSD(T)}] \approx 1.085\,(E[\text{CCSDT-3}] - E[\text{CCSD(T)}]) \\ - 1.515\,(E[\text{CCSDT-1b}] - E[\text{CCSD(T)}]) \\ + 0.200\,E[(T)] \quad (13)$$

This equation yields RMSD = 0.013 kcal/mol for CCSDT(Q)−CCSD(T), but system **22** is clearly an outlier as it was for the (Q) fit. Deleting **22** reduces RMSD to 0.009 kcal/mol; reoptimizing the parameters does not affect RMSD to the precision given, and the parameters change only minimally to {1.090, −1.502, 0.204}.

As a sanity check, we consider the recently reported [22] reduced-scaling CCSDT(Q) [43,44] calculations on the sandwich structure of naphthalene dimer. With the cc-pVDZ(d,p) basis set, these authors find a CCSDT(Q)-CCSD(T) difference of −0.67 kJ mol$^{-1}$ (−0.160 kcal mol$^{-1}$) for the naphthalene sandwich, and −0.32 kJ mol$^{-1}$ (−0.076 kcal mol$^{-1}$) for the corresponding benzene sandwich (which is well known [45] to be a saddle point). Using Eq. (12) we obtain estimates of −0.150 and −0.055 kcal mol$^{-1}$, respectively, while the corresponding values with Eq. (13) are −0.201 and −0.089 kcal/mol, respectively. The two sets of estimates closely bracket the actual calculated values, at a tiny fraction of the latter's computational expense.

A further transferability check is afforded by applying Eq. (12) to the subset of the S22 benchmark [46] for which we were able to complete the CCSDT(Q)/cc-pVDZ(d,s) calculations. At first sight, the RMSD doubled to 0.021 kcal/mol, but this was found to be entirely due to the formic acid dimer (**3**). In the absence of $p$ functions on the acidic hydrogens, one anomalously find $D_e[\text{CCSD}_{\text{corr}}] < D_e[(T)]$, which causes Eq. (12) to 'misbehave'. With any larger basis set (or for that matter, even for the larger acetic acid dimer **20** in S66), the problem disappears. Table 3 presents S66-optimized parameters with different basis sets used for the CCSD(T) calculations: as can be seen there, $\beta$ in particular is quite sensitive to the basis set. This can be attributed to $D_e[\text{CCSD}_{\text{corr}}]$ growing faster with the basis set than $D_e[(T)]$; see Ref. [10] and Refs. therein for a detailed analysis of the basis set convergence of (T).

In response to a reviewer comment, we considered a thermochemical application of Eq. (12), namely, to the W4-11nonMR ('non-multireference') subset of the W4-11 thermochemical benchmark [47], where 19 species with strong static correlation were omitted. (We also omitted the two beryllium halide species.) The post-CCSD(T) valence correlation contributions to the total atomization energies (TAEs) of these 122 species are 0.86 kcal mol$^{-1}$ RMS; applying Eq. (12) with the S66 fitted parameters $\alpha = 1.139$, $\beta = -0.273$, $a_3 = 1$ yields RMSD = 0.39 kcal mol$^{-1}$. Minimizing RMSD with respect to the three parameters brings this down to 0.27 kcal mol$^{-1}$ for $\alpha = 1.149$, $\beta = -0.163$, $a_3 = 0.322$. Eq. (5) yields RMSD = 0.25 kcal mol$^{-1}$ with $c_1 = 0.944$, $c_2 = -0.613$, and $c_3 = 0.213$. These statistics imply that, despite post-CCSD(T) corrections in small-molecule thermochemistry being at least an order of magnitude more important than in noncovalent interactions, Eqs. (12), (5) are still useful here as a semiquantitative estimate (and hence, as a fairly low-cost predictor of whether the more rigorous calculation will affect the species of interest significantly).

Finally, Table 4 presents a comparison between the present results and those obtained for nine complexes (six parallel-displaced $\pi$ stacks **24-29** and three T-shaped complexes **47-49**) by Grüneis and coworkers [15]. Clearly, their quasiperturbative CCSD(cT) approach yields higher-order triples contributions remarkably close to full CCSDT and actually outperforms CCSDT-3 in this regard. (One cannot help being reminded of how quasiperturbative CCSD(T) clearly outperforms its fully iterative counterpart [48] CCSDT-1.) The connected quadruples at first sight seem to 'reopen the gap' between CCSD(cT) and FN-DMC somewhat; however, the stochastic uncertainties on the FN-DMC results are so large that the noise overwhelms whatever signal there might be. Longer FN-DMC simulations with a smaller uncertainty would be highly desirable. It does however seem clear, from our results (particularly those from actual CCSDT and estimated (Q), rather than from actual CCSDT-3 and Eq. (12), that aromatic $\pi$ stacks are indeed overbound at the CCSD(T) level, albeit not by as much as CCSD(cT) and FN-DMC would seem to indicate.

## 4. Conclusions

We can spell out our conclusions here:

1. CCSD(T) and CCSD(T)$_\Lambda$ both benefit from error compensation and are both closer to CCSDT(Q) than either CCSDT-3 or full CCSDT. CCSD(T)$_\Lambda$ yields better results for $\pi$ stacks but slightly overbinds London complexes.
2. Already for small aromatic $\pi$-stacks in the S66 dataset, one observes nontrivial discrepancies between CCSD(T) and CCSDT(Q). This gap opens up further for naphthalene dimer, and can be expected to grow further for larger aromatic $\pi$-stacks.
3. The lion's share of $T_3-(T)$ is recovered at the CCSDT-3 level, and CCSDT-3 – CCSD(T) can be well estimated from (T) by a simple two-parameter formula, while the same is possible for CCSDT – CCSD(T) if CCSDT-1b and CCSDT-3 results are available.
4. The best predictor of (Q) is a scaled CCSDT-3 – CCSD(T) difference.
5. The CCSDT(Q)–CCSD(T) difference can actually be reproduced to about 0.01 kcal mol$^{-1}$ RMSD using a two-parameter formula, Eq. (12) with $a_3 = 1$, or a three-parameters formula, Eq. (13). Neither requires more strenuous calculations than CCSDT-3, and hence no steeper scaling with system size than $O(N^7)$

*Note added in revision*

As the revision of this manuscript was being finalized, a preprint reporting FN-DMC results for S66 came out [49]. That study appears to indicate that, directly opposite to its behavior for $\pi$ stacks, FN-DMC actually binds hydrogen-bonded dimers more *strongly* than CCSD(T). Reported discrepancies between FN-DMC and CCSD(T) reach as high as 0.7 kcal/mol for **20** AcOH... AcOH — these are very hard to rationalize as remaining flaws of CCSD(T), as Lao also very recently concluded [50].

*Note added in proof*

At page proof correction time, we additionally were able to converge the CCSDT/cc-pVDZ(d,s) total energy for species **17**, the Watson-Crick uracil dimer, to better than 0.001 kcal/mol. We can hence supplement the entry in Table 1 for this species with a CCSDT–CCSDT-3 difference of $-0.035$ kcal mol$^{-1}$. The latter implies a CCSDT–CCSD(T) difference of $-0.131$ kcal mol$^{-1}$, in excellent agreement with the estimate of $-0.127$ kcal mol$^{-1}$ obtained from Eq. (5).

**CRediT authorship contribution statement**

**Emmanouil Semidalas:** Writing – review & editing, Methodology, Investigation, Data curation, Conceptualization. **A. Daniel Boese:** Writing – review & editing, Methodology, Conceptualization. **Jan M.L. Martin:** Writing – review & editing, Writing – original draft, Resources, Project administration, Funding acquisition, Formal analysis, Data curation, Conceptualization.

**Declaration of competing interest**

The authors declare the following financial interests/personal relationships which may be considered as potential competing interests: Jan M.L. Martin reports financial support was provided by Israel Science Foundation. If there are other authors, they declare that they have no known competing financial interests or personal relationships that could have appeared to influence the work reported in this paper.

**Acknowledgments**

This work was supported by the Israel Science Foundation (grant 1969/20) and by the Uriel Arnon Memorial Center for AI research into smart materials. JMLM thanks the Quantum Theory Project at the University of Florida and its head, Prof. John F. Stanton, for their hospitality. The authors would like to thank Dr. Margarita Shepelenko (Weizmann) for critically reading the manuscript prior to submission, and also, together with Drs. Peter R. Franke (U. of Florida), Branko Ruscic (Argonne National Laboratory), Nisha Mehta (Weizmann) and Prof. Leslie Leiserowitz (Weizmann) for inspiring discussions. Special thanks are due to Dr. James R. Thorpe and Prof. Devin A. Matthews (Southern Methodist U., Dallas, TX) for fixing a particularly vexing memory allocation bug in CFOUR, and thus removing an insurmountable obstacle for the S66 calculations. All calculations were carried out on the ChemFarm HPC cluster of the Weizmann Institute Faculty of Chemistry.

**Appendix A. Supplementary data**

Microsoft Excel workbook with the relevant total and interaction energies.

Additional raw data may be obtained from the corresponding author upon reasonable request.

Supplementary material related to this article can be found online at https://doi.org/10.1016/j.cplett.2025.141874.

**Data availability**

Data shared as Supporting Information.